\documentstyle{mn}
\newcommand{\re}{\par\hangindent=0.5cm\hangafter=1\noindent}

\def\ltsima{$\; \buildrel < \over \sim \;$}
\def\ltsim{\lower.5ex\hbox{\ltsima}}
\def\gtsima{$\; \buildrel > \over \sim \;$}
\def\gtsim{\lower.5ex\hbox{\gtsima}}

\title{The origin of galactic disks with exponential z-profiles}

\author[A. Burkert and Y. Yoshii] 
       {A. Burkert$^{1}$ and Y. Yoshii$^{2,3,4}$\\
        $^1$Max-Planck-Institut f\"{u}r Astronomie, K\"onigstuhl 17
, D-69117, Heidelberg, Germany\\
        $^2$Institute of Astronomy, Faculty of Science,
University of Tokyo, Mitaka, Tokyo 181, Japan\\
        $^3$Research Center for the Early Universe, School of Science\\
          University of Tokyo, Bunkyo-ku, Tokyo 113, Japan\\
        $^4$Theoretical Astrophysics Center, Blegdamsvej 17 DK-2100 Copenhagen
\O,
Denmark}

\date{Received ...}

\begin{document}

\date{\small {\it Subject heading}: galaxies:spiral - galaxies:formation}

\maketitle

 \begin{abstract}
  A new solution is presented for the puzzling, observed universality
of the exponential luminosity profiles, perpendicular to the disk
plane of spiral and lenticular galaxies. It is shown that such exponential
$z$-profiles result naturally from gaseous protodisks which settle into
isothermal equilibrium prior to star formation. Subsequent cooling
leads to a gravitational contraction of the gas towards the equatorial
plane and to a stellar exponential $z$-profile if the
star formation rate is assumed to be comparable to the cooling rate of
the gas. The final stellar scale height depends only on the initial
gas temperature and local surface density. This model therefore provides
a new method to investigate the early energetic state of galactic
protodisks with measured scale heights and surface densities along
the disk plane.
 \end{abstract}

\section{Indroduction}

Galaxies of similar morphological type exhibit similarities in their 
structural properties, independent of whether they are isolated in 
general fields or agglomerated in galaxy clusters.  These similarities 
under different environments may emerge due to some {\it internal} 
regularities that have superseeded external, probabilistic disturbances 
such as interactions and/or mergers.

There is a well-known evidence from optical observations that many 
edge-on spiral and lenticular galaxies have a universal luminosity profile 
perpendicular to their disk plane, or a universal $z$-distribution of 
stellar mass density if a constant mass-to-luminosity ratio is assumed 
(Tsikoudi 1977; Burstein 1979; van der Kruit \& Searle 1981).  It was 
once claimed that the $z$-distribution is best fitted by a self-gravitating 
isothermal model like sech$^2(z/2h_z)$ using a scale parameter $h_z$ 
(van der Kruit \& Searle 1981). 

Near-infrared observations of edge-on spiral galaxies however uncovered an 
excess over the isothermal model at small $|z|$ where the optical photometry 
is hindered by dust absorption (Wainscoat, Freeman, \& Hyland 1989; 
Aoki et al. 1991; van Dokkum et al. 1994). These infrared observations 
showed that an exponential distribution $\exp(-|z|/h_z)$ provides a superior 
fit to the observed $z$-profiles.  Furthermore it is known from analyses of 
star counts that the vertical structure of the Galactic disk in the solar
neighbourhood is well fitted by an exponential distribution rather than an 
isothermal model (Bahcall \& Soneira 1980; Prichet 1983).  

An exponential $z$-distribution can be constructed by adding up several
stellar disk components with different vertical velocity dispersions, but 
this is only possible if the contribution from stars with larger velocity 
dispersion is fine-tuned to dominate progressively at larger distances from 
the disk plane.  A mechanism that enables this tuning has not been found 
to date.

The vertical disk structure has been
considered as a result of dynamical evolution of the stellar component through
encounters with massive clouds and spiral structures
(Villumsen 1983; Lacey 1984; Carlberg 1987). However, although collisions with clouds
might be important in increasing the velocity dispersions of young stars 
this effect cannot account for the high velocity dispersion of old stars 
(Lacey 1984). In addition, the resulting disks are isothermal and
not exponential.

We report that the hydrodynamical equations associated with a simple
star formation law possess a remarkable solution which naturally produces
an exponential stellar $z$-distribution in the final stage of gravitational
settling of galactic protodisks.  This result might provide new insight
into the formation of galactic disks and their early evolutionary phases.

\section{The numerical model}

The key factor which makes galaxy simulations distinct from either 
hydrodynamical computations is star formation.   From theoretical 
arguments (Hoyle 1953; Low \& Lynden-Bell 1976; Silk 1977), the cooling 
of the gas plays a decisive role in the process of star formation as it 
provides the required cool environments where density fluctuations in the 
gas can become gravitationally unstable on stellar mass scales.  It is 
therefore reasonable to assume that stars form at a rate comparable to the 
local cooling rate.  It is this working hypothesis that we use in this paper.

We will restrict ourselves to the gravitational settling of the protodisk
in the vertical direction, neglecting radial motions. This approximation
is reasonable because of the observational fact that galactic disks are in
general rotationally supported. The hydrodynamical equations which describe 
the evolution of the gas density $\rho$, the vertical gas velocity $w$ and 
the thermal gas energy by unit mass $e$ are then
\begin{eqnarray}
   \frac{\partial \rho}{\partial t} + \frac{\partial}{\partial z}
   (\rho w) & = & - \frac{\rho}{t_*} \nonumber \\
   \frac{\partial w}{\partial t} + w \frac{\partial w}{\partial z} +
   \frac{2}{3} \frac{\partial}{\rho \ \partial z} (\rho e) - g_z & = & 0 \\
   \frac{\partial e}{\partial t} + w \frac{\partial e}{\partial z} +
   \frac{2}{3} e \frac{\partial w}{\partial z} & = & - \frac{e}{t_c}
   \;\;\;, \nonumber
\end{eqnarray}
with
\begin{equation}
\partial g_z/\partial z=-4\pi G(\rho+\rho_*) \;\;\; ,
\end{equation}
where $\rho_*$ is the stellar density and the other notations have their usual
meanings.  The source terms on the right hand side of equation (1) describe 
the condensation of gas into stars on a timescale $t_*$ and the gaseous energy 
dissipation on a timescale $t_c=\rho e/\Lambda$ with the cooling rate per unit 
volume $\Lambda (\rho ,e)$. We introduce a free parameter $k=t_*/t_c$ which 
determines the timescale of star formation $t_*$ relative to $t_c$.   
Note that $k$ should be of order unity according to our assumption that star 
formation is related primarily to the cooling activity of the gas.

The cooling rate is a function of the local gas density and temperature and
therefore can be approximated as $\Lambda = \Lambda_0 \rho^a e^b$.  Typical 
values are $a=2$ and, in an ionized hydrogen gas, $b=+0.5$ or $b=-0.5$ for 
free-free cooling or free-bound cooling, respectively.  Note that possible 
heating would compete with the gas cooling and affect $a$ and $b$.  The 
power-law model also includes the possibility that energy dissipation in the 
disk is dominated by cloud-cloud collisions. In such a case $a=2$ and $b=1.5$ 
(Larson 1969).  In order to keep the simulation simple we treat $a$ and $b$ 
as free parameters and do not specify cooling and heating sources in an 
explicit way.

The stars will stay most of their time at maximum $z$-distance, that is
approximately at the position where they formed initially.  This is at least
valid for a majority of stars that make up the disk where the gas is quickly
converted into stars and subsequent contraction of the gas no longer has a 
significant dynamical effect on the stellar orbits 
(e.g., Yoshii \& Saio 1979).  
As a first approximation we therefore neglect the secular dynamical evolution
of stars and assume that their local density is given by
\begin{equation}
\rho_*(z) = \int_{0}^{\infty}\frac{\rho(z,t)}{t_*(z,t)} dt \ \  .
\end{equation}

A commonly perceived theoretical idea on the formation of galaxies
(Fall \& Efstathiou 1980; Blumenthal et al. 1986; Barnes \& Efstathiou 1987)
is that their initial state is more or less the virialized isothermal sphere
where baryonic matter relaxes with dark matter.  A more recent idea from
cosmological simulations (Katz 1992) is that spiral galaxies formed through 
the hierarchical merging of smaller subunits that were disrupted in the 
merging process. Whereas their dissipationless stellar and dark matter 
components now constitute the visible and dark halos of disk galaxies, the 
gas could dissipate its kinetic energy, spin up and settle into the equatorial
plane where it formed an initially extended, gaseous protodisk.  As we will 
demonstrate below, the observed exponential vertical disk profiles indicate 
that this protodisk was able to achieve a local isothermal equilibrium 
state prior to the onset of star formation.  The initial disk temperature 
$T(r)$ then is only a function of the disk radius $r$ but independent of $z$, 
and the initial surface density $\Sigma (r)$ determines the initial central 
density $\rho (z=0)$ at radius $r$ and its local scale height $h_z (r)$:

\begin{equation}
\rho(z)=\frac{\Sigma}{4h_z}{\rm sech}^2\left(\frac{z}{2h_z}\right) \;\;,
\;\;\;\;\;
h_z=\frac{k_B T}{2\pi G\Sigma\mu m_H} \;\; ,
\end{equation}
where $\mu$ is the mean molecular weight.  We use this as a reasonable initial
setup in our calculations.


The hydrodynamical equations are solved on an Eulerian staggered grid,
with 100 logarithmically spaced grid points in $z$ and an outer edge placed
at a sufficiently large $z$-distance.  The set of differential equations
is integrated numerically by means of an explicit, finite difference scheme
with operator splitting and monotonic transport.   This scheme provides
second-order accuracy in space and time (Burkert \& Hensler 1987).
The simplicity of the above equations allows us to make a comprehensive survey 
in the parameter space.

\section{Discussion of the numerical results}

A number of test calculations have shown that the final stellar $z$-profile
depends strongly on the chosen initial distribution of the protodisk gas if 
one starts from a non-equilibrium state.  On the other hand, an exponential 
stellar disk forms in all cases where we assume the gas to settle into 
isothermal equilibrium, prior to star formation and gas cooling, as long as 
the star formation time $t_*$ is comparable to the cooling time $t_c$, that 
is, $k\sim 1$.

In the limit of rapid star formation ($k\gg 1$) the initial isothermal
distribution of the gas is frozen in the stellar disk which has more or 
less a flat-top $z$-profile near the equatorial plane. In the limit of 
slow star formation ($k\ll 1$) the gas condenses into the equatorial 
plane towards the energy minimum state, giving rise to a power-law 
$z$-profile in the stellar disk.  Exponential profiles are in between 
these two extremes, and our calculations show that when $k$ lies inside 
a preferable range $0.3\ltsim k\ltsim 3$, the final stellar $z$-profile 
becomes exponential, independent of the free parameters $k$, $\Lambda_0$, 
$a$ and $b$, and also independent of $T$ and $\Sigma$ of the isothermal 
protodisk gas.  This remarkable result still holds when the isothermal 
equilibrium state is modified to some extent.

In Figure 1 we show as an example a sequence of models with changing $b$
(upper panel) or changing $a$ (lower panel) while the other parameters are
fixed as shown in the panels: $k=1$, $T=10^5$K, $\Sigma=50M_\odot$pc$^{-2}$,
and $\Lambda_0$ set to give $\langle t_c\rangle =10^9$yrs at the half-mass
$z$-distance of the protodisk.  The stellar $z$-profile $\rho_*(z)$ is almost 
perfectly exponential all the way from $z\sim0$ to 10$h_z$ spanning about 
five orders in density, and the scale length becomes systematically smaller 
for larger $b$ or larger $a$.  It is evident that the final scale height 
depends critically on the adopted values of the parameters and in general 
differs from the initial scale height of the gaseous protodisk.

In Figure 2 we show one of the other sequences of models with decreasing
$\Sigma$ from the left to the right, using $a=2$, $b=0$ and the other 
parameters fixed as in Figure 1.  $\Sigma$ is chosen to mimic the situation 
in our own Galaxy with a surface density distribution of 
$\Sigma=50M_{\odot}$pc$^{-2}\times \exp[-(R-8$kpc)/4kpc]
evaluated at galactocentric radial intervals $R=2$kpc$\times i$ with 
$i=1$ to 5.  Note that the profiles have been shifted along the horizontal 
axis.  The stellar $z$-profile $\rho_*(z)$ is almost perfectly exponential
for different choices of $\Sigma$, and the stellar disk flares with
decreasing surface density, that is, its thickness increases as 
$h_z\propto\Sigma^{-1}$.  Such a $\Sigma$-dependence of $h_z$ would be 
expected if the stellar system remembers the exponential tail of the 
self-gravitating isothermal disk model which was adopted as an initial 
condition (see Eq.4).  Note however that the final exponential stellar
density distribution extends much further inwards, till $z=0$.
Increasing $b$ while keeping $\Sigma $ and the other parameters constant
decreases the scale height according to $h_z(b=1) = 0.6 \times h_z(b=0)$
and $h_z(b=2) = 0.4 \times h_z(b=0)$.

The disk flaring, when seen edge-on, could largely be masked by projection
effects.  In Figure 3 we show the edge-on projected $z$-profiles $\Sigma_*(z)$ 
at the same radii as in Figure 2 along the disk plane from the galaxy center.  
The projection makes the profiles less dependent on $R$.  In such a case 
the inversion for recovering the original density profiles $\rho_*(z)$ is 
generally very unstable.  A great accuracy is needed in measuring 
$\Sigma_*(z)$ over a wide range of $R$ and $z$, in order to detect the disk 
flaring from the observations of edge-on spiral galaxies.

While the resulting stellar $z$-profiles are always exponential, the scale
height $h_z$ itself depends on the cooling rate $\Lambda (\rho ,e)$ through 
$a$ and $b$ and also on the initial state of the protodisk gas through $T$ 
and $\Sigma$.  In practice, however, the deterministic quantities are only 
$T$ and $\Sigma$ because if these are given the cooling source is virtually 
specified.  Then, $a$ and $b$ are are no longer free parameters.  In other 
words, given that $h_z$ is measured at sufficient accuracy along the disk 
plane, we could constrain $T$ and $\Sigma$ as a function of radial distance
from the galaxy center.  This result can be used in order to explore the
global initial structure of the protodisk from the vertical disk structure 
of an edge-on galaxy observed today.

\section{Conclusions}
 
The presented idea that the star formation timescale $t_*$ is comparable
to the cooling timescale $t_c$ for the origin of exponential stellar 
$z$-profiles works when the protodisk has nearly reached an equilibrium 
prior to star formation and gas cooling. 

Under a realistic situation the dynamical timescale $t_{dyn}$ of the 
protodisk may initially not be in balance with the cooling timescale $t_c$.  
If $t_{dyn}<t_c$ in the hot and tenuous gas, the protodisk 
achieves an equilibrium state and undergoes a
quasi-static contraction until the gas density becomes sufficiently 
large so that $t_c$ falls below $t_{dyn}$.  If $t_{dyn}>t_c$ otherwise, 
the gas cools rapidly and condenses into cool and dense clouds.  In this 
stage, cloud collisions dissipate kinetic energy and enhance the
formation of stars. As demonstrated by Burkert et al. (1992), the increase in
energy input through supernova explosions will
compensate energy dissipation and cooling, 
eventually halting the collapse and achieving a quasi-equilibrium state.
In either cases, after reaching its 
equilibrium, the protodisk evolves due to the feedback 
effects of star formation, independent of its initial formation history
and of variations in global physical conditions (e.g., Lin \& Murray 1992).

Gas cooling or energy dissipation induces a slow gravitational settling 
of the protodisk while, at the same time, leading to the formation of stars in 
the disk.  This coupling yields a continuous change of various stellar 
characteristics (age, metallicity, color, velocity dispersion, etc) as a 
function of the $z$-distance from the disk plane.  Although our model is 
not sufficiently detailed to allow an extensive test against such data, 
the power of producing an exponential stellar $z$-profile from a vast range 
of physical conditions is very appealing and should be considered as a strong 
candidate for understanding the universality of exponential $z$-profiles.

Burkert et al. (1992) have performed more detailed simulations of the
dynamical settling of hot protogalactic gas into the  equatorial plane, taking 
into account star formation and heating and cooling processes in a multiphase 
interstellar medium.  They achieved a good agreement with the observed 
vertical density structure of the Galactic disk in the solar neighborhood.  
The complexity of their model leads however to a very large parameter space 
which cannot be explored in detail due to computational limitations.   
Among many different processes involved in their physical model, the process 
of crucial importance is that the rate of star formation is 
adjusted sooner or later to balance with the local cooling rate by means of
the self-regulated star formation mechanism (Cox 1983; Franco \& Cox 1983).  
It is not clear from their models, which process is most
important in leading to exponential $z$-profiles.
Our calculations demonstrate that an ideal condition for such a profile
is  $t_*\sim t_c$, independent of the details of heating and cooling.

\subsection*{Acknowledgments}

We thank K. Freeman and H. Saio for invaluable comments on the subject
discussed in this paper.   A.B. acknowledges the financial support from 
the Program of Human Capital Mobility, Grant Number ERBCHGECT920009 and 
of CESCA Consortium and the hospitality of the Centre d'Estudis Avancats 
de Blanes.  Y.Y. acknowledges the financial support from the Yamada Science
Foundation, the German Academic Exchange Service (DAAD), and also Theoretical
Astrophysics Center, Denmark, under which this work was performed.  
This work has been supported in part by the Grant-in-Aid for COE Research 
(07CE2002).  All calculations have been performed on the Cray YMP 4/64 of 
the Rechenzentrum Garching.

\newpage

\begin{center}
{\small REFERENCES}
\end{center}

\re
\ Aoki, T. E., Hiromoto, N., Takami, H., \& Okamura, S.  1991, PASJ,
43, 755

\re
\ Bahcall, J. N., \& Soneira, R. M.  1980, ApJS, 44, 73

\re
\ Barnes, J., \& Efstathiou, G.  1987, ApJ, 319, 575

\re
\  Blumenthal, G. R., Faber, S. M., Flores, R. A., \& Primack, J. R.
1986, ApJ, 301, 27

\re
\ Burkert, A., \& Hensler, G.  1987, MNRAS, 208, 493

\re
\ Burkert, A., Truran, J., \& Hensler, G. 1992, ApJ, 391, 651

\re
\ Burstein, D.  1979, ApJ, 234, 829

\re
\ Carlberg, R. G.  1987, ApJ, 322, 59

\re
\ Cox, D. P.  1983, ApJL, 265, 61

\re
\ Fall, S. M., \& Efstathiou, G.  1980, MNRAS, 193, 189

\re
\ Franco, J., \& Cox, D. P.  1983, ApJ, 273, 243


\re
\ Hoyle, F.  1953, ApJ, 118, 513

\re
\ Katz, N. 1992, ApJ, 391, 502

\re
\ Larson, R.B. 1969, MNRAS, 145, 405

\re
\ Lacey, C. G. 1984, MNRAS, 208, 687

\re
\ Lin, D. N. C., \& Murray, S. D.  1992, ApJ, 394, 523

\re
\ Low, C., \& Lynden-Bell, D.  1976, MNRAS, 176, 367

\re
\ Pritchet, C.  1983, AJ, 88, 1476

\re
\ Silk, J.  1977, ApJ, 214, 152

\re
\ Tsikoudi, V.  1977, Univ. Texas Publ. Astron. No. 10

\re
\ van der Kruit, P. C., \& Searle, L. 1981, A\&A, 105, 115

\re
\ van Dokkum, P. G., Petetier, R. F., de Grijs, R., \& Balcells, M.  1994,
A\&A, 286, 415

\re
\ Villumsen, J. V. 1983, ApJ, 274, 632 

\re
\ Yoshii, Y., \& Saio, H.  1979, PASJ, 339, 368

\re
\ Wainscoat, R. J., Freeman, K. C., \& Hyland, A.R.  1989, ApJ, 337, 163

\clearpage

\begin{center}
{\small FIGURE CAPTIONS}
\end{center}

\vspace{1.0cm}

{\normalsize F{\footnotesize IG}.}
1.---The final stellar $z$-profile $\rho_*(z)$ for different values of the
power index $b$ (upper panel) or $a$ (lower panel) for the cooling rate defined
as $\Lambda = \Lambda_0 \rho^a e^b$, with other parameters fixed in the model,
as shown in the figure (for details see text).
The $z$-profiles (thick lines) originate from the same initial
distribution
of the protodisk gas (dotted line).

\vspace{1.0cm}

{\normalsize F{\footnotesize IG}.}
2.---The final stellar $z$-profile $\rho_*(z)$ for different values
of the initial surface mass density of the protodisk $\Sigma$ (face-on
projected), with other parameters fixed in the model (for details see text).
The five $z$-profiles from the left to the right show a sequence of
decreasing $\Sigma$, according to the formula appropriate to our own Galaxy:
$\Sigma=50M_{\odot}$pc$^{-2}\times \exp[-(R-8$kpc)/4kpc] evaluated at
galactrocentric radial intervals $R=2$kpc$\times i$ with $i=1$ to 5.
Note that the profiles have been shifted along the horizontal axis.

\vspace{1.0cm}

{\normalsize F{\footnotesize IG}.}
3.---The edge-on projected $z$-profiles $\Sigma_*(z)$ at the same radii as
in Figure 2 along the disk plane from the galaxy center.
Note that the profiles have been shifted along the horizontal axis.

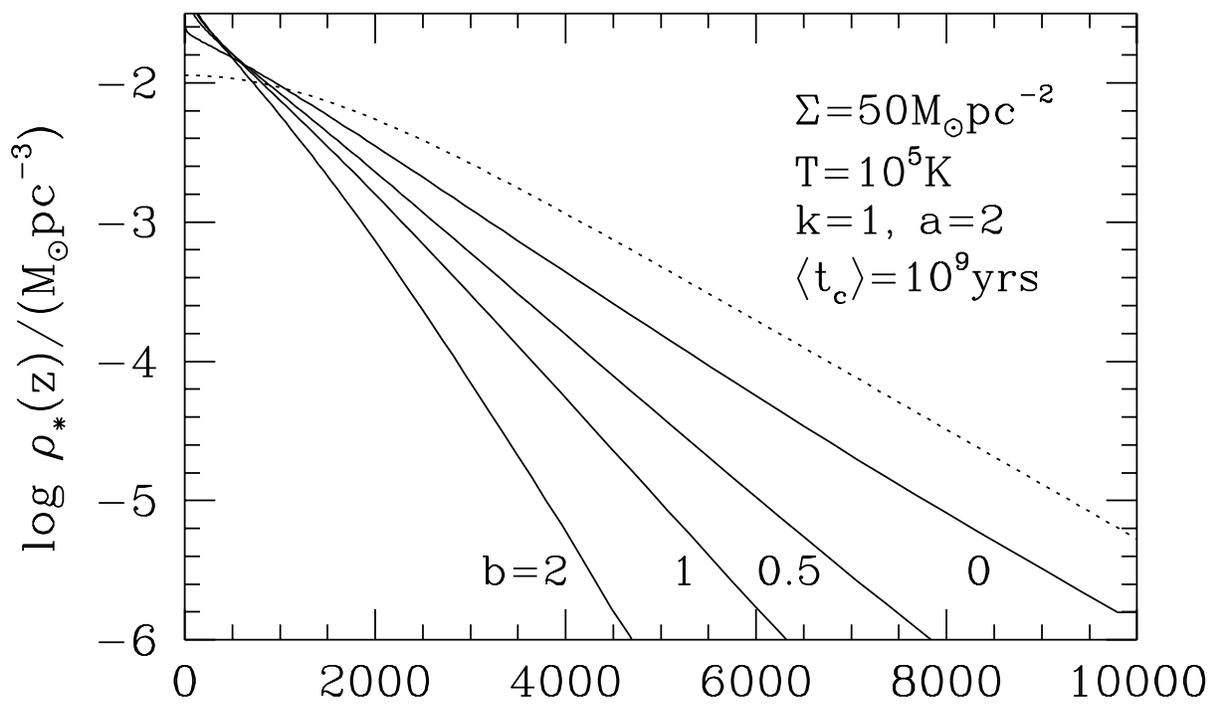
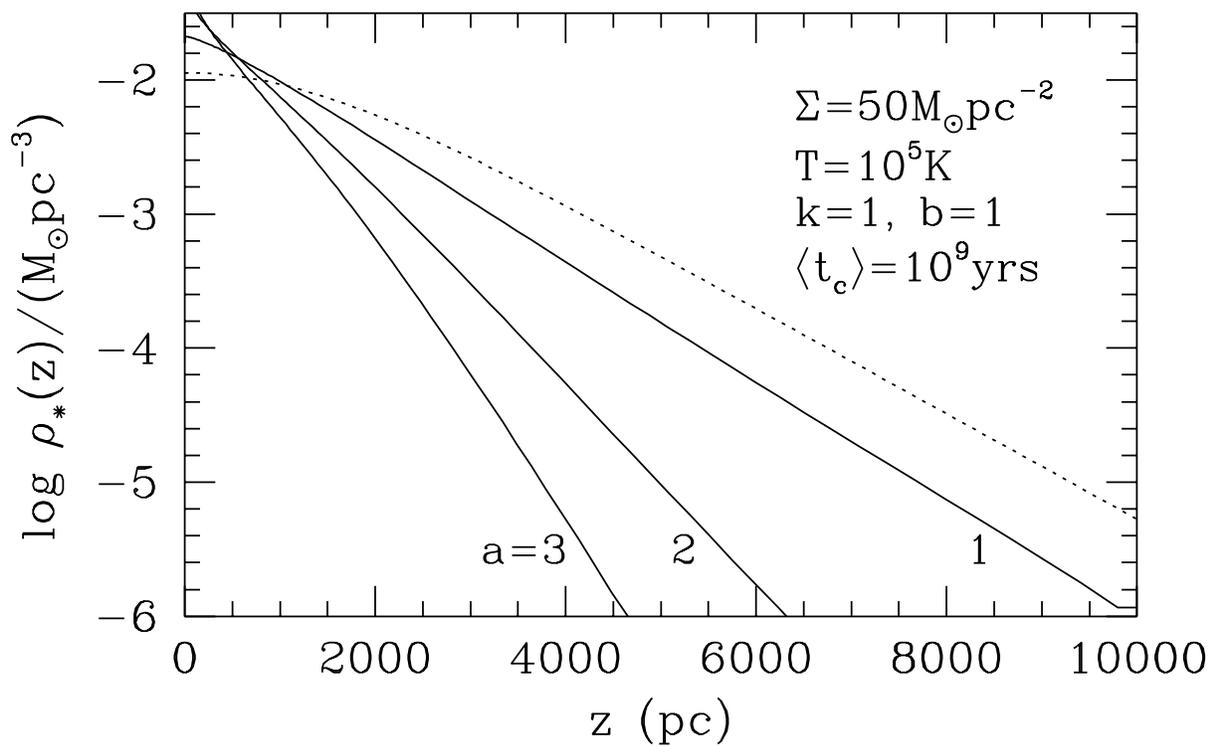
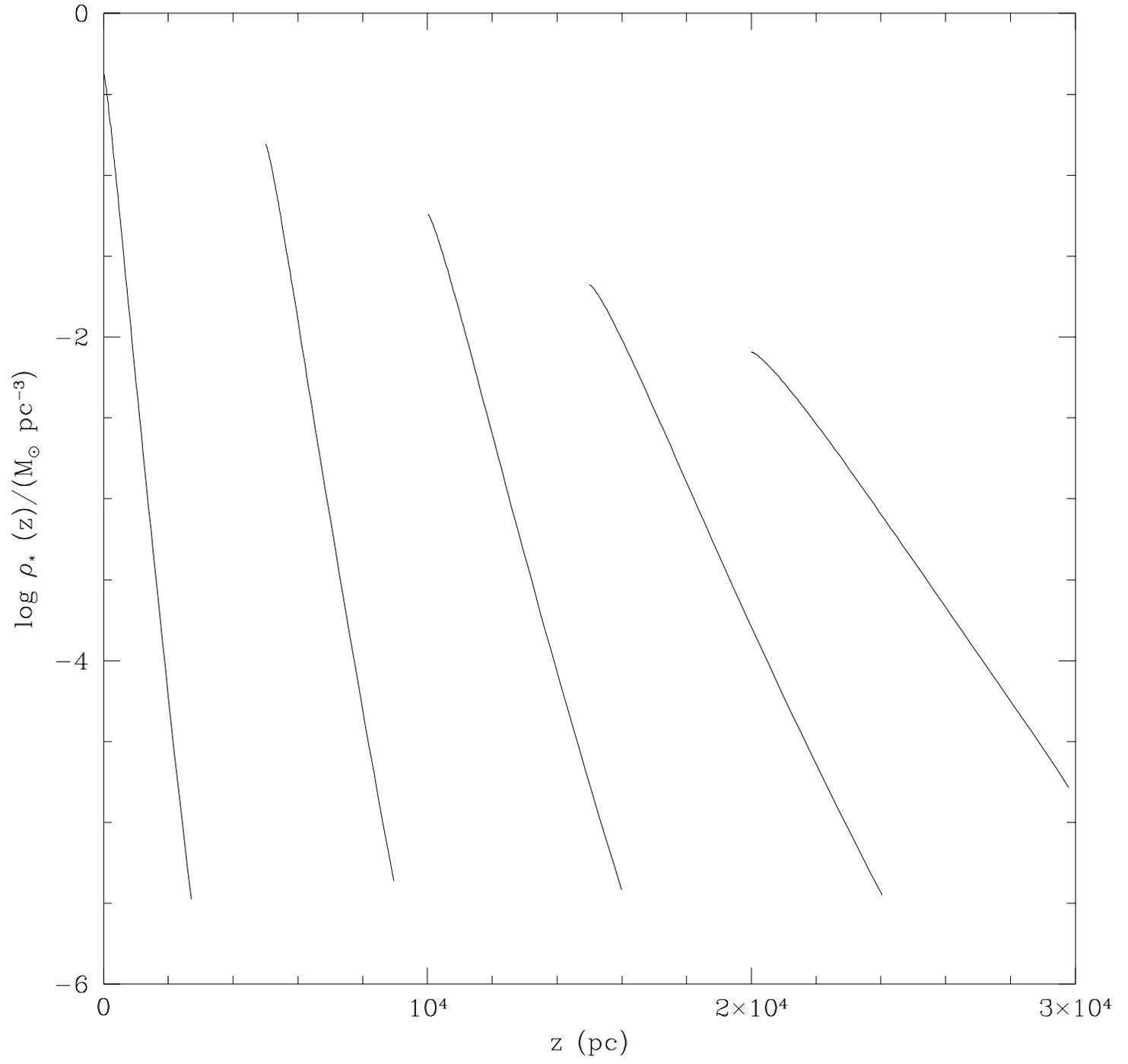
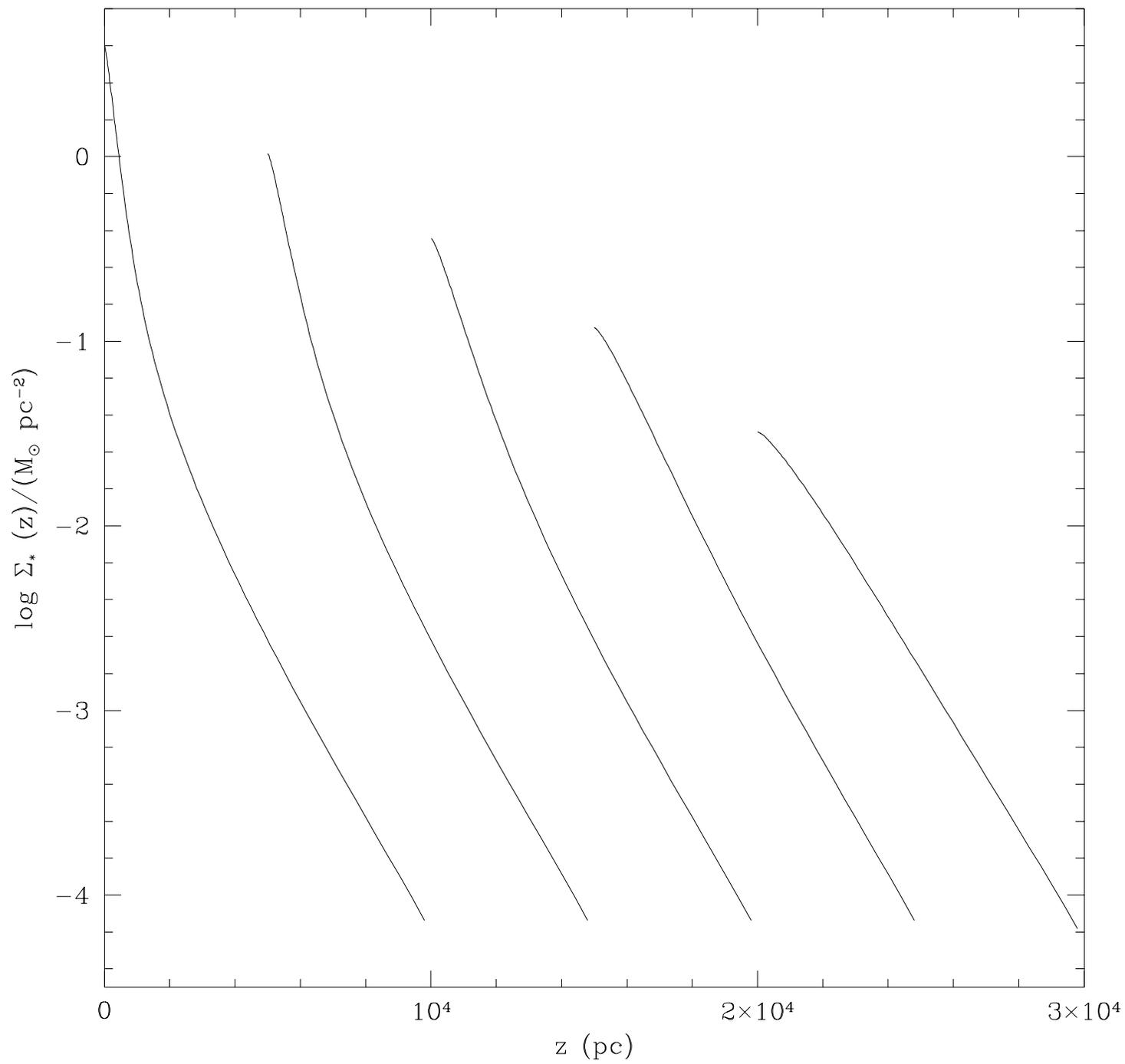
\end{document}